# AUTHENTICATION AND AUTHORIZATION IN MICROSERVICE-BASED SYSTEMS: SURVEY OF ARCHITECTURE PATTERNS

Alexander Barabanov[1], Denis Makrushin[2]

**Abstract**

**Context.** Service-oriented architecture and its microservice-based approach increase an attack surface of applications. Exposed microservices become a pivot point for advanced persistent threats and completely change the threat landscape. Correctly implemented authentication and authorization architecture patterns are basis of any software maturity program.
**Objective**. The aim of this study is to provide a helpful resource to application security architect and developers on existing architecture patterns to implement authentication and authorization in microservices-based systems.
**Method**. In this paper, we conduct a review of major electronic databases and libraries as well as security standards and presentations at the major security conferences.
**Results**. In this work based on research papers and major security conferences presentations analysis we identified industry best practices in authentication and authorization patterns and its applicability depending on environment characteristic. For each described patterns we reviewed its advantages and disadvantages that could be used as decision-making criteria for application security architects during architecture design phase.

## 1 Introduction

The microservice architecture is being increasingly used for designing and implementing application systems in both cloud-based and on-premise infrastructures, high-scale applications and services [1]. There are many security challenges need to be addressed in the application design and implementation phases. The fundamental security requirements that have to be addressed during design phase are authentication and authorization. Therefore it is vital for applications security architects to understand and properly use existing architecture patterns to implement authentication and authorization in microservices-based systems. The goal of our research was to identify such patterns and to do recommendations for applications security architect on possible way to use it. This study is conducted with three main questions in mind:

- Which architecture patterns to implement authentication and authorization have been reported in microservice-based systems researches?
- What advantages and disadvantages do existing architecture patterns have?
- What should application security architect take in mind while selecting pattern to implement authentication and authorization in microservice-based systems?

We reviewed major electronic databases and libraries (IEEE Xplorer, ACM Digital Library, SpringerLink) with research papers to extract primary studies. In order to explore these databases and presentations, we used search strings containing "authentication", "authorization", "service-oriented architecture" and "microservice" (in different spelling, like "micro-service" or "micro service") words. To avoid missing relevant studies, we also reviewed security standards, presentations at the major security conferences.

In summary, this paper makes the following contributions:

- a state of the art of the authentication and authorization architecture patterns for microservice-based systems (Section 2);
- recommendations for applications security architect on how to select an appropriated architecture pattern (Section 3).

---

[1] Alexander Barabanov Ph.D, CISSP, CSSLP Advanced Software Technology Laboratory, Huawei.
[2] Denis Makrushin, OSCP, Advanced Software Technology Laboratory, Huawei.

# 2 Authentication and authorization architecture patterns

We made decomposition of authentication and authorization functions based on microservice-specific characteristics and identified the following list of security sub-functions (Figure 1): edge-level authorization, service-level authorization, external entity identity propagation and service-to-service authentication. Then we reviewed major electronic databases and libraries as well as security standards and presentations at the major security conferences in order to identify existing architectural patterns.

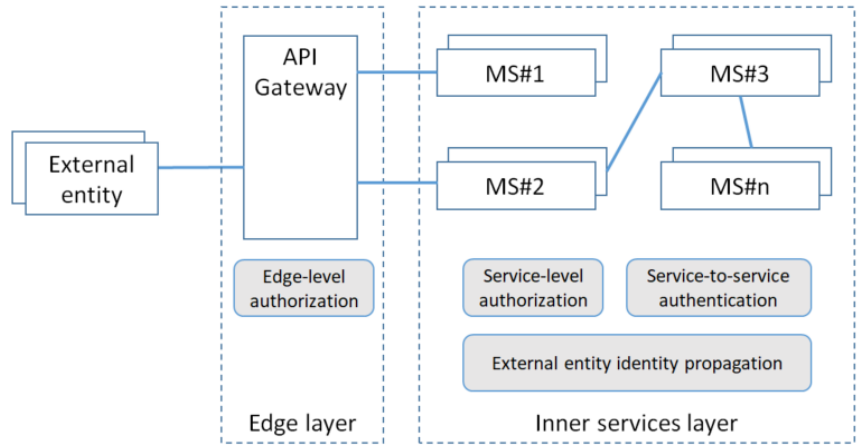

Figure 1 Authentication and authorizationsubfunctions in microservice-based systems

## 2.1 Edge-level authorization

In simple scenario authorization can happen only at the edge level (API gateway). The API gateway can be leveraged to centralize enforcement of authorization for all downstream microservices, eliminating the need to provide authentication and access control for each of the individual services [2]. In such case, NIST recommends [3] to implement mitigating controls such as mutual authentication to prevent direct, anonymous connections to the internal services (API gateway bypass). It should be noted that authorization at the edge layer has a following drawbacks [4]:

- pushing all authorization decisions to API gateway can quickly become hard to manage in complex ecosystems with many roles and access control rules;
- API gateway may become a single-point-of-decision that may violate "defense in depth" principle;
- operation teams typically own the API gateway, so development teams can not directly make authorization changes, slowing down velocity due to the additional communication and process overhead.

In most cases, development teams implement authorization in both places - at the edge level at a coarse level of granularity and service level [5]. To authenticate external entity edge can use access tokens (referenced token or self-contained token) transmitted via HTTP headers (e.g. "Cookie" or "Authorization") or use mTLS [6].

## 2.2 Service-level authorization

For further discussion, we use terms and definitions (Figure 2) according with NIST [7]. The functional components of access control system can be classified following way:

- Policy Administration Point (PAP) provides a user interface for creating, managing, testing, and debugging access control rules;
- Policy Decision Point (PDP) computes access decisions by evaluating the applicable access control policy;

- Policy Enforcement Point (PEP) enforces policy decisions in response to a request from a subject requesting access to a protected object;
- Policy Information Point (PIP) serves as the retrieval source of attributes, or the data required for policy evaluation to provide the information needed by the PDP to make the decisions.

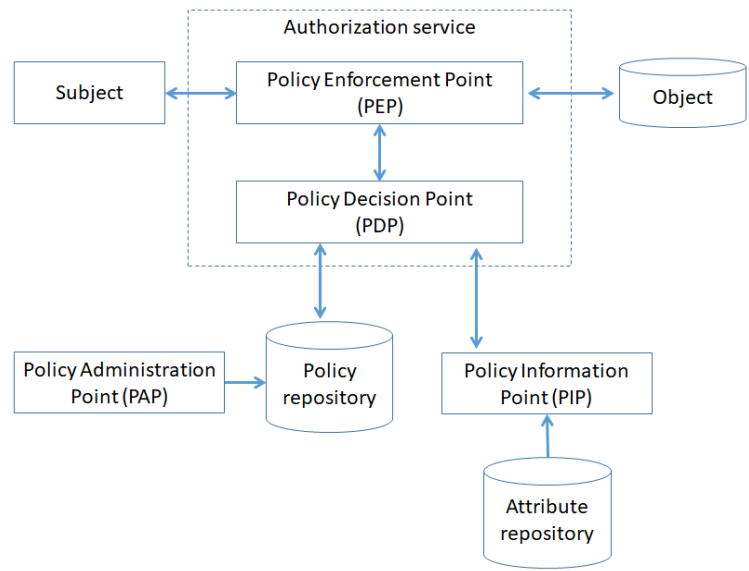

Figure 2 Access control management functional points [7]

Service-level authorization gives each microservice more control to enforce access control policies. Based on review of [4, 6, 8, 9] we identified the following types of service-level authorization:

- decentralized pattern;
- centralized pattern with a single Policy Decision Point (PDP);
- centralized pattern with an embedded PDP.

### 2.2.1 Decentralized pattern

In that pattern development team implements PDP and PEP directly at microservice code level (Figure 3). All the access control rules and as well as attributes that need to implement that rule are defined and stored on the each microservice (step 1). When microservice receives (step 2) request along with some authorization metadata (e.g., end-user context), microservice analyzes it (step 3) in order to generate access control policy decision and then enforces authorization (step 4).

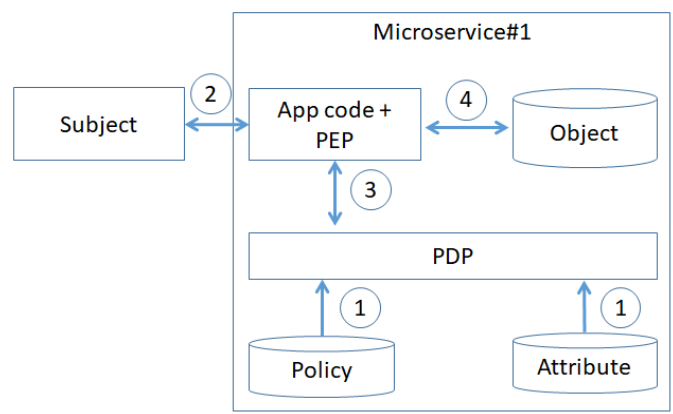

Figure 3 Decentralized pattern high-level architecture

Existing programming language frameworks allow development teams to implement authorization at the microservice layer. Implementing authorization at the source code level means that the code must be updated whenever development team want to modify authorization logic and has following limitation [4, 10]:

- each development team must clearly understand security features of using programming language framework and implement its correctly in their microservices;
- each development team must clearly understand access control policy and expected permissions for a role/group that may be challenging task because the decisions are potentially littered through one or more large, complicated code bases;
- this pattern relies on the careful manual configuration by the development team, which is error-prone; besides that, due to the large scale of modern microservice applications, it is unrealistic for the development team to configure and maintain access control policies for every microservice;
- source code changes require solid regression testing for authorization bugs detection.

On the other hand implementing access control policy in the microservice code allows developers to enforce more fine granting access control because rules that govern authorization are more domain specific [8].

### 2.2.2 Centralized pattern with single policy decision point

In that pattern access control rules are defined, stored, and evaluated centrally (Figure 4). Access control rules is defined using PAP (step 1) and delivered to centralized PDP as well as attributes that need to implement that rules (step 2). When a subject invokes microservice endpoint (step 3), microservice code invokes centralized PDP via network call and PDP generates access control policy decision by evaluating the query input against access control rules and attributes (step 4). Based on PDP decision microservice enforce authorization (step 5).

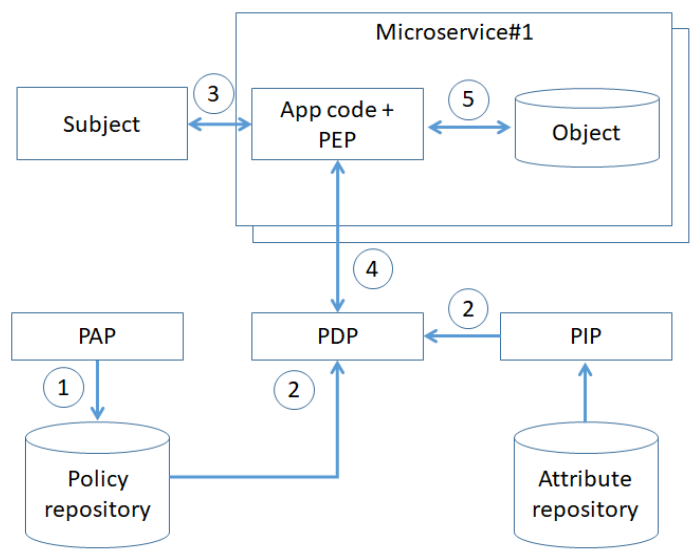

Figure 4 Centralized pattern with single PDP high-level architecture

Several benefits of this pattern are [4, 11]:

- security/development team can update access control rules without changing the source code, it enables centralized policy management, changes in policy may be deployed separately from microservices using them;
- access control rule definitions can be left for development teams to implement, but left outside of the core business portion of source code to make policies discoverable, moreover some access control rules are "universal" and may be shared within a microservices/organization;

- access control rules may be used in operations environments to detect security anomalies, e.g. anomaly microservice behavior based on API call, on threat detection it is possible to dynamically re-create and apply new access control rules to mitigate security risk.

To define access control rules development/operation team has to use some language or notation. An example is Extensible Access Control Markup Language (XACML) and Next Generation Access Control (NGAC) that is a standard to implement policy rules description [12, 13]. However, XACML ended up failing because it required learning a separate, complicated syntax, causing more work for developers, and there were not many open source integrations [4].

This pattern badly affects latency due additional network calls of the remote PDP endpoint, but it can be mitigated by caching authorization policy decisions at microservice level [6]. It should be mentioned that PDP must be operated in high-availability mode due to resilience and availability issues. Application security architects should combine it with other patterns (e.g., authorization on API gateway level) in order to avoid "single-point-of-decision" and enforce "defense in depth" principle.

### 2.2.3 Centralized pattern with embedded policy decision point

In that pattern access control rules are defined centrally but stored and evaluated at microservice level (Figure 5). Access control rules is defined using PAP (step 1) and delivered to embedded PDP as well as attributes that need to implement that rules (step 2). When a subject invokes microservice endpoint (step 3), microservice code invokes PDP and PDP generates access control policy decision by evaluating the query input against access control rules and attributes (step 4). Based on PDP decision microservice enforce authorization (step 5) [14].

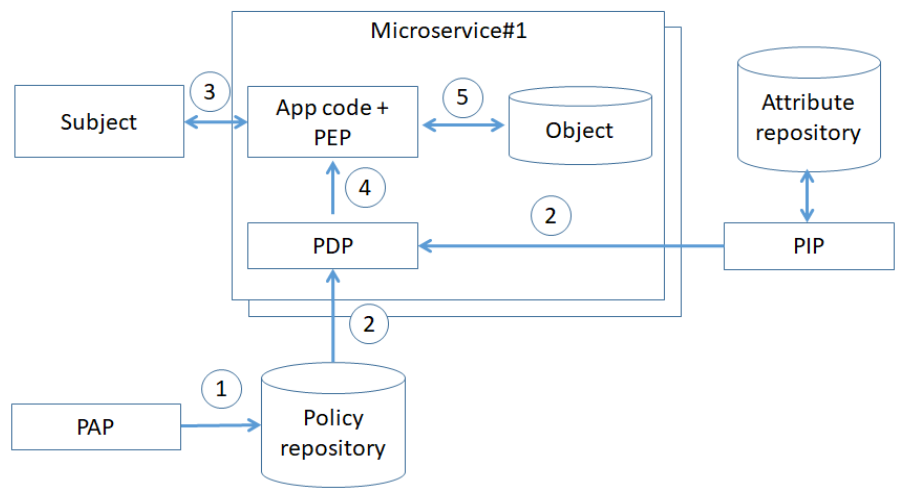

Figure 5 Centralized pattern with embedded PDP high-level architecture

PDP code in that case can be implemented as microservice built-in library or sidecar in service mesh architecture [15-22]. Due to possible network/host failures and network latency it is advisable to implement embedded PDP as microservice library or sidecar on the same host with microservice. Embedded PDP usually store authorization policy and policy-related data in-memory to minimize external dependencies during authorization enforcement and get low latency [23]. Main difference from "Centralized pattern with single policy decision point" with caching approach is that authorization decisions do not store on the microservice side, up to date authorization policy are stored on microservice side instead. It should be mentioned that caching authorization decisions may lead to applying outdated authorization rules and access control violations.

M.Mehta and T. Sandall presented [23, 24] a real case of using "Centralized pattern with embedded PDP" pattern to implement authorization on the microservices level (Figure 6):

- Policy portal and Policy repository is UI-based system for creating, managing and versioning access control rules;
- Aggregator fetches data used in access control rules from all external sources and keeps it up to date;
- Distributor pulls access control rules (from Policy repository) and data used in access control rules (from Aggregators) to distribute it among PDPs;
- PDP (library) asynchronically pulls access control rules and data and keeps it up to date to enforce authorization by PEP component.

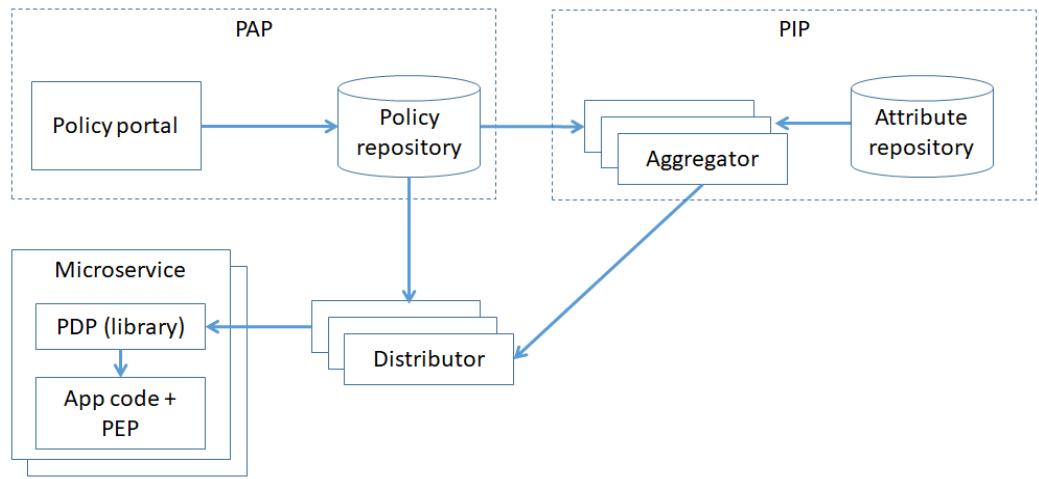

Figure 6 Centralized pattern with embedded PDP (example)

Benefits of this patterns are the same as for "Centralized pattern with single PDP" plus pattern does not badly affect latency due to embedding PDP on the microservice level.

There are several challenges that have to be taken into account while applying this pattern:

- this pattern relies on the manual or semi-manual access policy rules designed by security team that may be error-prone – security testing and verification practices should be implemented to avoid insecure configuration vulnerabilities;
- application security architects should combine it with other patterns (e.g., authorization on the edge level) in order to avoid "single-point-of-decision" and enforce "defense in depth" principle;
- it may be the case that some business-specific access control rules cannot be implemented in that way – application security architects should combine that pattern with "Decentralized pattern";
- application security architects should choose an approach of how to get authorization policy updates from the centralized PAP (e.g., PAP polling or publish-subscribe mechanism [6]):
- development team has to securely use 3$^{rd}$–party authorization components and describe access control policy using some formal language that in some cases may be overhead – "Decentralized pattern" may be enough to implement some simple access control policy.

## 2.3. External entity identity propagation

To make fine-granted authorization decision at the microservice level microservice has to understand caller context (e.g. user ID, user roles/groups). In order to allow internal service layer to enforce authorization edge layer has to propagate authenticated external entity identity (e.g., end-user context) along with a request to downstream microservices. One of the simplest way to propagate external entity identity is to re-use the access token received by the edge and pass it to internal microservices. It should be mentioned that approach is highly insecure due to possible external access token leakage and may decrease an attack surface because the communication relies on proprietary token-based system implementation and internal microservices have to understand external access token [25]. This pattern also is not external access token agnostic, i.e. internal services have to support a wide range of authentication techniques to extract identity from different types of external tokens (e.g. JWT, cookie, OpenID Connect token).

There are two patterns to pass the external entity identity from one microservice to another [6, 26]:

- send the external entity identity as a clear or self-signed data structures;
- send the external entity identity as a data structures signed by the trusted issuer.

### 2.3.1 Send the external entity identity as a clear or self-signed data structures

In that approach calling microservice extracts external entity identity from incoming request (e.g. via parsing incoming access token), creates data structure (e.g. JSON or self-signed JWT) with context and passes that on to an internal microservices [27] (Figure 7).

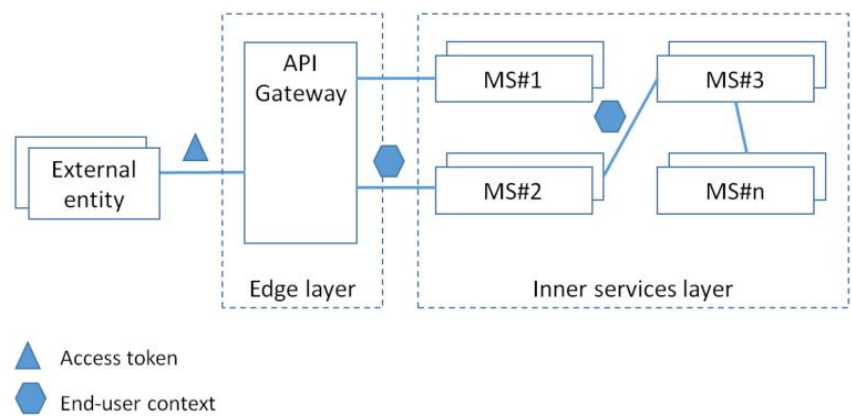

Figure 7 Pattern "Send the user context as a clear or self-signed data structures"

In this scenario recipient microservice has to trust the calling microservice - if the calling microservice want to violate access control rules, it can do so by setting any user/client ID or user roles it wants as the HTTP header [6]. That approach is applicable in a highly trusted environment in which every microservice is developed by trusted development team according with secure software development practices [28, 43].

### 2.3.2 Using a data structures signed by a trusted issuer

In this pattern after the external request is authenticated by authentication service at the edge layer, a data structure representing external entity identity (e.g., contained user ID, user roles/groups or permissions) is generated, signed or encrypted by the trusted issuer and propagated to internal microservices [6, 25, 26, 29]. S. Thadeshwar [30] presented a real case of using that pattern: structure called "Passport" that contains user ID and its attributes and HMAC protected is created at the edge level for each incoming request, propagated to internal microservices and never exposes outside (Figure 8):

1. Edge authentication service (EAS) obtains secret key from the Key Management System.
2. EAS receives an access token (may be e.g. in a cookie, JWT, OAuth2 token) from incoming request.
3. EAS decrypts the access token, resolves the external entity identity and sends it to the internal services in the signed "Passport" structure.
4. Internal services can extract user identity in order to enforce authorization (e.g. to implement identity-based authorization) using wrappers.
5. If necessary, internal service can propagate "Passport" structure to downstream services in the call chain.

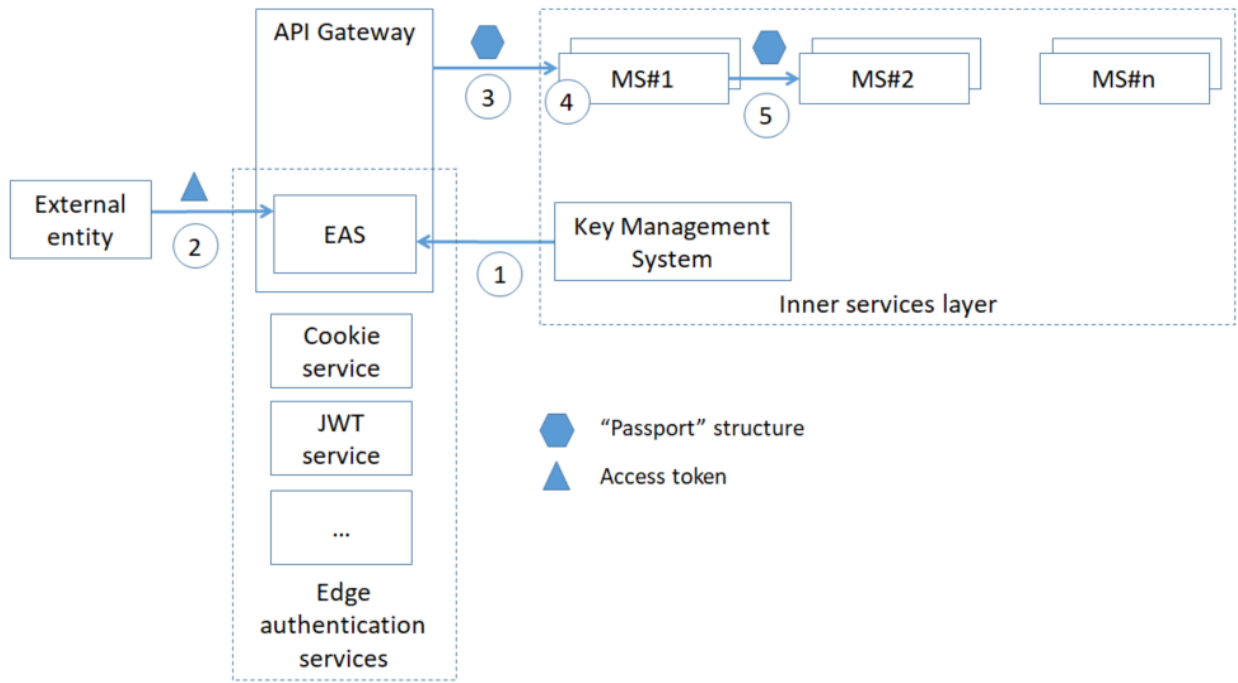

Figure 8 Using a data structures signed by a trusted issuer (example)

It should be mentioned that pattern is external access token agnostic and allows to decouple external entity and its internal representation.

## 2.4 Service-to-service authentication

There are two common ways to implement service-to-service authentication [6]:

- mutual transport layer security (mTLS);
- token based, e.g. JSON Web Tokens (JWT).

In mTLS approach each microservice can legitimately identify who it talks to, in addition to achieving confidentiality and integrity of the transmitted data. Each microservice in the deployment has to carry a public/private key pair and uses that key pair to authenticate to the recipient microservices via mTLS. mTLS usually is implemented with a self-hosted Public Key Infrastructure [6, 26]. The main challenges using mTLS are: key provisioning and trust bootstrap, certificate revocation and key rotation.

Token based approach works at the application layer. Token is a container and may contain caller ID (microservice ID) and its permissions (scopes). Caller microservice can obtain token by invoking special security token service using its own service ID and password and then attaches it to every outgoing requests e.g., via HTTP headers. In most cases, token-based authentication works over TLS that provides confidentiality and integrity of data in transit [6].

According to [6, 26, 33-36] mTLS is the most popular option to authenticate microservices.

Network segmentation or firewalling pattern implements "trust-the-network" approach in which no security is enforced in service-to-service communication and currently not widely used by community as primary security mechanism [6, 31].

## 3. Recommendations for application security architects

Based on our survey results, we came up with several recommendations for application security architects on authentication and authorization implementation.
Summary on authorization implementation is presented in the table below (Table 1).

| Pattern name | Scalability | Latency | 3rd-party components dependencies | Reconfiguration on-the-fly |
|---|---|---|---|---|
| Decentralized pattern | Low | Low | not in use | Not supported |
| Centralized pattern with single policy decision point | High | High | in use | Supported |
| Centralized pattern with embedded policy decision point | High | Low | in use | Supported |

Recommendation on how to implement authorization are the following.
1) To achieve scalability it is not advisable to hardcode authorization policy in source code (decentralized pattern), but use special language to express policy instead. The goal is to externalize/decouple authorization from code, and not just with a gateway/proxy that acts as a checkpoints. Recommended pattern for service-level authorization is "Centralized pattern with embedded PDP" due to its resilience and wide adoption.
2) Authorization solution should be platform-level solution; dedicated team (e.g., Platform security team) must be accountable for development and operation of authorization solution as well as sharing microservice blueprint/library/components that implement authorization among development teams.
3) Authorization solution should be based on widely used solution, because implementing custom solution has following cons [4]:
    - security or engineering team have to build and maintain custom solution;
    - it is necessary to build and maintain client library SDKs for every language used in system architecture;
    - necessity to train every developer on custom authorization service API and integration, and there's no open source community to source information from.
4) There is a probability that not all access control policy can be enforced by gateways/proxies and shared authorization library/components, so some specific access control rules still have to be implemented on microservice buisnes code level. In order to do that it is advisiable to have and use by microservice development teams simple questionary/check-list to uncover such security requirments and handle its properly during microservice development.
5) It is advisable to implement "defense in depth" principle - enforce authorization on:
    - gateways and proxies level at a coarse level of granularity;
    - microservice level using shared authorization library/components to enforce fine-granted decisions;
    - microservice business code level to implement business-specific access control rules.
6) Access control policy formal procedures like development, approvement, rolling-out must be implemented.

Summary on external entity identity propagation is presented in the table below (Table 2).

Table 2

| Pattern name | Applied environment | 3rd-party components dependencies | External access token agnostic | Ability to centralize |
|---|---|---|---|---|
| Send the external entity identity as a clear or self-signed data structures | Trusted | not in use | non-agnostic | - |
| Using a data structures signed by an trusted issuer | Untrusted | In use | agnostic | + |

Recommendation on how to propagate external entity identity among microservices are the following.
1) In order to implement external access token agnostic and extendable system decouple access tokens issued for external entity from its internal representation. Use single data structure to represent and propagate external entity identity among microservices. Edge-level service has to verify incoming external access token, issue internal entity representation structure and propagate it to downstream services.
2) Using an internal entity representation structure signed (symmetric or asymmetric encryption) by a trusted issuer is recommended pattern adopted by community [30].
3) Internal entity representation structure should be extensible to enable add more claims that may lead to low latency.
4) Internal entity representation structure must not be exposed outside (e.g., to browser or external device).

Recommendations on how to proper implement authentication in microservice-based systems are the following.
1) mTLS is widely used and recommended pattern to implement service-to-service authentication.
2) mTLS solution should be platform-level solution based on widely used solution, because implementing proprietary solution has following cons:
    a. in-house development must be highly experienced in cryptography in order to implement it in a right way;
    b. in-house development team will have to build, maintain it and fix vulnerabilities;
    c. it is necessary to train every developer on custom authentication service API and integration, and there's no open source community to source information from.
3) dedicated team (e.g., Platform security team) must be accountable for development and operation of mTLS solution as well as sharing among development teams microservice blueprint/library/components that implement it.
4) It is advisable to implement "defense in depth" principle, e.g. enforce authentication using network segmentation or firewalling as a secondary security control.

## 4. Related work

Security architecture patterns for microservice-based systems has been the topic of a number of surveys and review articles, as well as standards.

Vale et al. [37] conducted a systematic mapping to reveal adopted security mechanisms for microservice-based systems. They focused only on security mechanisms and examined 26 papers published from November 2018 to March 2019.

Hannousse et al. [38] conducted a similar investigation to Vale et al. [37] study. Their study is broader in several ways: they included published papers since 2011 and besides security mechanisms, they also focused on identifying security threats and the applicability of proposed solutions regarding their execution platforms and architectural layers.

Yu et al. [39] surveyed work related to security risks for microservices-based fog applications, and argued that security issues arise in four system aspects: containers, data, permissions and network security.

NIST published [3, 21] a standards on microservice-based system security. NIST analyzed the multiple implementation options available for each individual core security feature (authentication and access management, service discovery, secure communication protocols, security monitoring, availability/resiliency improvement techniques, load balancing and throttling, integrity assurance techniques and handling of session persistence) and configuration options in architectural frameworks, and developed security strategies that counter threats specific to microservice-based systems.

Several research papers propose authorization schemes and policies related to microservice-based systems. Fu et al. [40] by studying the traditional access control technologies derived its limitations and shortcomings in the microservice environment, and proposed an access control optimization model based on role based access control (RBAC). Triartono et al. [41] proposed a model to implement RBAC on OAuth 2.0 using Laravel framework. Liu et al. [42] introduced security boundary of basic platform, business system and system function in the intelligent campus, defined a hybrid access control strategy based on group based access control, RBAC and hierarchical policy-based access control models. Those works mainly focus on access control policy formal design and verification and do not pay much attention to design/architecture issues.

Compared with the related works our study is more narrow and concentrated on authentication and authorization only in order to get deeper results. Moreover, besides research papers analysis we also analyzed presentations at the major security conferences.

# 5. Conclusion and further work

Correctly implemented authentication and authorization functions are the basis for further step of microservice-based infrastructure hardening and software maturity program improvement. In the survey, we identified industry best practices in authentication and authorization architecture patterns, its advantages and disadvantages and its applicability depending on environment characteristic. For each described patterns we reviewed its advantages and disadvantages that could be used as decision-making criteria for security architects, considering authentication and authorization implementation in service-oriented environment.

Microservices creates new security challenges:

- increase attack surface of modern applications;
- decrease effectiveness of traditional logging systems that relies on centric-based log-aggregation architecture;
- blur of development lifecycle across multiple components of application instead on monolithic application;
- increase traffic level due to growing number of communications between microservices.

The challenges requires novel methods of monitoring and threat detection even based machine learning techniques [44, 45] that take into account the specificity of microservice operations.

# References

[1] A. Boubendir, E. Bertin and N. Simoni, "A VNF-as-a-service design through micro-services disassembling the IMS," 2017 20th Conference on Innovations in Clouds, Internet and Networks (ICIN), Paris, 2017, pp. 203-210, doi: 10.1109/ICIN.2017.7899412.

[2] D. Lu, D. Huang, A. Walenstein and D. Medhi, "A Secure Microservice Framework for IoT," 2017 IEEE Symposium on Service-Oriented System Engineering (SOSE), San Francisco, CA, 2017, pp. 9-18, doi: 10.1109/SOSE.2017.27.

[3] Chandramouli R. (2019) Security Strategies for Microservices-based Application Systems. (National Institute of Standards and Technology, Gaithersburg, MD), NIST Special Publication (SP) 800-204. https://doi.org/10.6028/NIST.SP.800-204

[4] Lakshminarayanan S. (2019). Authorization in Micro Services World Kubernetes, Istio and Open Policy Agent. Talk presented at the AppSecCali 2019

[5] Stivers C., Higgins N. (2019). Deploying Open Policy Agent at Atlassian. Talk presented at the OPA Summit 2019

[6] Microservices Security in Action, Prabath Siriwardena and Nuwan Dias, 2020, Manning

[7] Vincent C. Hu, Ferraiolo D., Kuhn R., Schnitzer A., Sandlin K., Miller R., Scarfone K. (2014) Guide to Attribute Based Access Control (ABAC) Definition and Considerations. (National Institute of Standards and Technology, Gaithersburg, MD), NIST Special Publication (SP) 800-162. https://doi.org/10.6028/NIST.SP.800-162

[8] Karthik K. (2018). Microservices Identity & Authorization. Talk presented at the XCon 2018

[9] Grandja J. (2019). Implementing Microservices Security Patterns & Protocols with Spring Security. Talk presented at the Spring I/O 2019

[10] Li, Xing & Chen, Yan & Lin, Zhiqiang. (2019). Towards Automated Inter-Service Authorization for Microservice Applications. SIGCOMM Posters and Demos '19: Proceedings of the ACM SIGCOMM 2019 Conference Posters and Demos. 3-5. 10.1145/3342280.3342288.

[11] Eknert A. (2019). Securing APIs with Open Policy Agent. Talk presented at the 2019 Platform Summit

[12] Nehme, Antonio & Jesus, Vitor & Mahbub, Khaled & Abdallah, Ali. (2019). Fine-Grained Access Control for Microservices. 10.1007/978-3-030-18419-3_19.

[13] David Ferraiolo, Ramaswamy Chandramouli, Rick Kuhn, and Vincent Hu. 2016. Extensible Access Control Markup Language (XACML) and Next Generation Access Control (NGAC). In Proceedings of the 2016 ACM International Workshop on Attribute Based Access Control (ABAC '16). Association for Computing Machinery, New York, NY, USA, 13–24. DOI:https://doi.org/10.1145/2875491.2875496

[14] D. Preuveneers and W. Joosen, "Towards Multi-party Policy-based Access Control in Federations of Cloud and Edge Microservices," 2019 IEEE European Symposium on Security and Privacy Workshops (EuroS&PW), Stockholm, Sweden, 2019, pp. 29-38, doi: 10.1109/EuroSPW.2019.00010.

[15] Sorens M. (2019). Open Policy Agent in Practice: From Angular to OPA in Chef Automate. Talk presented at the OPA Summit 2019

[16] Ray J. (2019). Open Policy Agent for Policy-enabled Kubernetes and CI/CD. Talk presented at the OPA Summit 2019

[17] Massa L. (2019). TripAdvisor: Building a Testing Framework for Integrating Open Policy Agent into Kubernetes. Talk presented at the OPA Summit 2019

[18] Huang Z. (2019). Deep Dive: Kubernetes Policy WG. Talk presented at the OPA Summit 2019

[19] Tao L. (2019). How We Use Istio and OPA for Authorization. Talk presented at the KubeCon + CloudNativeCon 2019

[20] Krach J., Fu W. (2019). Open Policy Agent at Scale: How Pinterest Manages Policy Distribution. Talk presented at the OPA Summit 2019

[21] Chandramouli R., Butcher Z. (2020) Building Secure Microservices-based Applications Using Service-Mesh Architecture. (National Institute of Standards and Technology, Gaithersburg, MD), NIST Special Publication (SP) 800-204A. https://doi.org/10.6028/NIST.SP.800-204A

[22] Rushgrove G.(2019). Applying Policy Throughout The Application Lifecycle with Open Policy Agent. Talk presented at the KubeCon + CloudNativeCon 2019

[23] Mehta M., Sandall T. (2018). The distributed authorization system: A Netflix case study. Talk presented at the Velocity Conference - San Jose, CA 2018

[24] Mehta M., Sandall T. (2017). How Netflix Is Solving Authorization Across Their Cloud. Talk presented at the KubeCon + CloudNativeCon 2017

[25] Brady S., Delegation Patterns for OAuth 2.0. Available at https://www.scottbrady91.com/OAuth/Delegation-Patterns-for-OAuth-20

[26] T. Yarygina and A. H. Bagge, "Overcoming Security Challenges in Microservice Architectures," 2018 IEEE Symposium on Service-Oriented System Engineering (SOSE), Bamberg, 2018, pp. 11-20.

[27] A. Bánáti, E. Kail, K. Karóczkai and M. Kozlovszky, "Authentication and authorization orchestrator for microservice-based software architectures," 2018 41st International Convention on Information and Communication Technology, Electronics and Microelectronics (MIPRO), Opatija, 2018, pp. 1180-1184, doi: 10.23919/MIPRO.2018.8400214.

[28] Alexander Barabanov, Alexey Markov, Andrey Fadin, Valentin Tsirlov, and Igor Shakhalov. 2015. Synthesis of secure software development controls. In Proceedings of the 8th International Conference on Security of Information and Networks (SIN '15). Association for Computing Machinery, New York, NY, USA, 93–97. DOI:https://doi.org/10.1145/2799979.2799998

[29] Ideskog J. (2016). Decoupling user identities from API design. Talk presented at the Nordic APIs Stack Event 2016.

[30] Thadeshwar S. (2019) User & Device Identity for Microservices @ Netflix Scale. Talk presented at the QCon 2019

[31] M. Pahl and L. Donini, "Securing IoT microservices with certificates," NOMS 2018 - 2018 IEEE/IFIP Network Operations and Management Symposium, Taipei, 2018, pp. 1-5, doi: 10.1109/NOMS.2018.8406189.

[32] Siriwardena P. (2020) Securing APIs with Transport Layer Security (TLS). In: Advanced API Security. Apress, Berkeley, CA

[33] Yung-Kao Hsu and S. P. Seymour, "An intranet security framework based on short-lived certificates," in IEEE Internet Computing, vol. 2, no. 2, pp. 73-79, March-April 1998, doi: 10.1109/4236.670687.

[34] B. Payne. (2016). PKI at scale using short-lived certificates. Talk presented at the USENIX Enigma, 2016

[35] Behrens S., Kanekar E. (2019). A Pragmatic Approach for Internal Security Partnerships. Talk presented at the AppSecCali 2019

[36] Wardrop M. (2019). Container Security: Theory & Practice at Netflix. Talk presented at the Docker Con 19

[37] A. Pereira-Vale, G. Márquez, H. Astudillo and E. B. Fernandez, "Security Mechanisms Used in Microservices-Based Systems: A Systematic Mapping," 2019 XLV Latin American Computing Conference (CLEI), Panama, Panama, 2019, pp. 01-10, doi: 10.1109/CLEI47609.2019.235060.

[38] Abdelhakim Hannousse, Salima Yahiouche. Securing Microservices and Microservice Architectures: A Systematic Mapping Study. URL: https://arxiv.org/abs/2003.07262

[39] Dongjin Yu, Yike Jin, Yuqun Zhang, and Xi Zheng. A survey on security issues in services communication of microservices-enabled fog applications. Concurrency and Computation: Practice and Experience, 31(22):e4436, 2019. e4436 cpe.4436.

[40] G. Fu, J. Sun and J. Zhao, "An optimized control access mechanism based on micro-service architecture," 2018 2nd IEEE Conference on Energy Internet and Energy System Integration (EI2), Beijing, 2018, pp. 1-5, doi: 10.1109/EI2.2018.8582628.

[41] Z. Triartono, R. M. Negara and Sussi, "Implementation of Role-Based Access Control on OAuth 2.0 as Authentication and Authorization System," 2019 6th International Conference on Electrical Engineering, Computer Science and Informatics (EECSI), Bandung, Indonesia, 2019, pp. 259-263, doi: 10.23919/EECSI48112.2019.8977061.

[42] B. Liu, Y. Yang and Z. Zhou, "Research on Hybrid Access Control Strategy for Smart Campus Platform," 2018 IEEE 3rd Advanced Information Technology, Electronic and Automation Control Conference (IAEAC), Chongqing, 2018, pp. 342-346, doi: 10.1109/IAEAC.2018.8577828.

[43] A. Barabanov, A. Markov and V. Tsirlov, "Procedure for substantiated development of measures to design secure software for automated process control systems," 2016 International Siberian Conference on Control and Communications (SIBCON), Moscow, 2016, pp. 1-4, doi: 10.1109/SIBCON.2016.7491660.

[44] Gaifulina D.A., Kotenko I.V. Application of deep learning methods in cybersecurity tasks. Voprosy Kiberbezopasnosti. №3(37), 2020. p. 76-86. (in Russ.) DOI: 10.21681/2311-3456-2020-03-76-86

[45] Sheluhin O.I., Ryabinin V.S., Farmakovskiy M.A., Anomaly detection in computer system by intellectictual analysis of system journals, Voprosy kiberbezopasnosti, №2(26), 2018. p 36-41. (in Russ.) DOI: 10.21681/2311-3456-2018-2-33-43